# Localised excitation of a single photon source by a nanowaveguide

*Wei Geng[1], Mathieu Manceau[2], Nancy Rahbany[1], Vincent Sallet[3], Massimo De Vittorio[4,5], Luigi Carbone[5], Quentin Glorieux[2], Alberto Bramati[2], Christophe Couteau[1,6]\**

[1] Laboratory of Nanotechnology, Instrumentation and Optics (LNIO), Charles Delaunay Institute, CNRS UMR 6281, University of Technology of Troyes (UTT), 10000, Troyes, France.

[2] Laboratoire Kastler Brossel, UPMC-Sorbonne Universités, CNRS, ENS-PSL Research University, Collège de France, 4 place Jussieu Case 74, F-75005 Paris, France.

[3] Groupe d'étude de la matière condensée (GEMAC), CNRS, University of Versailles St Quentin, 78035 Versailles Cedex, France.

[4] Istituto Italiano di Tecnologia (IIT), Center for Bio-Molecular Nanotechnologies Via Barsanti sn, 73010 Arnesano (Lecce), Italy.

[5] National Nanotechnology Laboratory (NNL), CNR Istituto Nanoscienze, Via per Arnesano km 5, 73100 Lecce, Italy.

[6] CINTRA CNRS-Thales-NTU UMI 3288, and School of Electrical and Electronic Engineering, Nanyang Technological University, 637553 Singapore.

*Correspondence and requests for materials should be addressed to C.C. (email: christophe.couteau@utt.fr).

**Abstract**: Nowadays, integrated photonics is a key technology in quantum information processing (QIP) but achieving all-optical buses for quantum networks with efficient integration of single photon emitters remains a challenge. Photonic crystals and cavities are good candidates but do not tackle how to effectively address a nanoscale emitter. Using a nanowire nanowaveguide, we realise an hybrid nanodevice which locally excites a single photon source (SPS). The nanowire acts as a passive or active sub-wavelength waveguide to excite the quantum emitter. Our results show that localised excitation of a SPS is possible and is compared with free-space excitation. Our proof of principle experiment presents an absolute addressing efficiency $\eta_a$ ~ $10^{-4}$ only ~50% lower than the one using free-space optics. This important step demonstrates that sufficient guided light in a nanowaveguide made of a semiconductor nanowire is achievable to excite a single photon source. We accomplish a hybrid system offering great potentials for electrically driven SPSs and efficient single photon collection and detection, opening the way for optimum absorption/emission of nanoscale emitters. We also discuss how to improve the addressing efficiency of a dipolar nanoscale emitter with our system.

Recently, the development of integration of nanoscale photonic elements has opened new horizons for information and communication processing and in particular for QIP[1]. Since the concept of quantum information processing was proposed using single photon sources[2] (SPSs), we witnessed the emergence of new platforms, such as photonic crystals[3–5], microcavities[6,7] or light resonators[8] where efficient light extraction is usually sought for. On the other hand, the efficient addressing and the optimisation of the non-resonant excitation of single emitters are eluded in such systems. A standard way of operating in order to excite a nanoscale emitter is to increase the excitation power until efficient absorption occurs. In any reasonable foreseen application, the amount of energy put in such a system will also have to be taken into careful consideration when nanophotonics will become ubiquitous to communication and information technologies. Due to their capacity to emit single photons on demand[1], albeit not easily addressable, zero-dimensional (0D) nanostructured materials are being considered as prospective basic components for quantum information processors. In the meantime, semiconductor nanowires (NW), synthesized by top-down or bottom-up methods, are being applied in nanophotonics as linking blocks for optical circuits due to their high-quality sub-wavelength waveguiding features [9–11]. As a consequence of their unique one-dimensional (1D) structure, they exhibit fascinating optical, electronic, and mechanical properties, which enable them for various NW-based devices such as photodetectors, nano-lasers and transistors[12–15]. The integration of 0D and 1D systems has attracted increasing interests in the emerging field of quantum photonics, as a 1D NW can favour the excitation, extraction and transfer of single photons emitted by a 0D nanostructure[16–19]. The so-called photonic wire is a good example of such system[20]. However, the fabrication of such devices usually involves complex fabrication steps.

We propose in this work a new hybrid system made of a nanocrystal and a 1D (a nanowire) nanostructure. We report the fabrication of a single photon source excited via a single semiconductor nanowire and we tackle the issue of the localised excitation of a single nanoemitter. The question of optimised absorption and extinction of emitters in general has been studied on single molecules[21,22,23] and single epitaxial quantum dots[24] using bulky microscopy techniques. In these studies, it was shown that an optical microscope can focused the light enough so that the beam size is on the order of the absorption cross-section of the emitter, given by $\sigma = 3\lambda^2/2\pi$ when excited at resonance. Aside the fact that these optical techniques cannot be integrated, it is known that the spatial mode profile of a nanoemitter can be quite complex[25], thus requiring specific care if one wants to obtain maximum optimum absorption. Photonics structures using thin films have also been investigated for efficient excitation [26]. The semiconductor nanowire in our system is made to operate as a passive or an active sub-wavelength nanowaveguide depending on whether we use a coupling light above or below its bandgap energy. Relative to the characteristic free-space excitation, we manage to measure with our set-up an addressing efficiency as large as 3 % which proves the possibility to excite a single nanoemitter using solely an excitation light coupled into a nanowaveguide. We measure the absolute addressing efficiency of our hybrid system to be $\eta_a \sim 0.6 \ 10^{-4}$ (see supporting information for more details) which is only 46 % smaller than the absolute addressing efficiency using our free-space. The measured addressing efficiency is corroborated by the value found using FDTD simulations (see supporting information).We can also show using FDTD simulation and with a resonant pumping that one can increase such an efficiency (see supplementary information) by carefully engineering this hybrid system. This will certainly open a new way for quantum integrated optics and complement existing techniques in nanophotonics with single emitters.

Our structure consists of a high-index semiconductor ZnO NW (inset Figure 1-a) drop-cast onto a quartz substrate and employed as a nanowaveguide to direct the excitation light towards a nanoscale emitter, the latter consisting of a CdSe/CdS core/shell nanocrystal (NC). Figure 1-a presents a schematic of the system showing a ZnO nano-waveguide exciting a visible-emitting single NC. ZnO NWs (3.3 eV energy bandgap) are grown by metal organic chemical vapour deposition (MOCVD). High-crystalline nanowires with fairly uniform diameters and defect-free surfaces[30] can be fabricated upon using this technique. Due to their high refractive index ($n_{ZnO}$ = 2.2 to 2.5 for the 370 to 410 nm range)[31], ZnO NWs are ideal candidates for light confinement and guiding within a broad wavelength range[32], which includes its own near-UV photoluminescence (PL) bandgap around 380 nm. Their dimensions can be controlled during the fabrication process and the NWs described in this article are 7~8 μm in length and 260~300 nm in diameter, as evidenced by scanning electron microscopy (SEM) shown in the inset of Figure 1-a.

Highly fluorescent elongated CdSe/CdS colloidal NCs are used as room temperature single photon sources[33,34]. Synthesized in liquid phase following a seeded growth approach[28], such NCs are made up by a CdSe spherical core of 2.7 nm in diameter, surrounded by an elongated shell of CdS thus exhibiting a final rod shape with dimensions of 50 nm in length and 7 nm in diameter[34,35]. CdSe/CdS NCs display a high PL quantum efficiency of 73 % measured with our set-up together with a good single photon emission[34].

A detailed description of the hybrid nano-device fabrication can be found in Supporting Information (SI), Figure S2. Briefly, ZnO NWs are deposited from a solution onto a PMMA-coated quartz plate. Then, PMMA stripes alternatively containing or not containing NCs are designed on top of the dispersed individual NWs using electron beam lithography (EBL). Figure 1-b presents a confocal reflection image of a single ZnO nanowire crossing the border between two PMMA stripes respectively embodying and free of NCs. By filtering out the excitation illumination light (455 nm LED), we can locate a single CdSe/CdS NC (bright dot) at one extremity of the NW (see Figure 1-c). This NW-NC hybrid system will be specifically studied in this article (another example of such system can be found in supplementary information with Figure S3).

Imaging and micro-PL spectroscopy are performed with a home-built confocal microscope combined with a high-resolution spectrometer. Thanks to the confocal configuration we have access to the spatial and spectral information of the PL signals. For imaging, we use an oil-immersion objective (100X, NA = 1.2) and a Peltier-cooled CCD camera at −80°C. Two continuous-wave lasers are used for the waveguiding of the NWs, respectively below ($\lambda_{exc}$ = 405 nm, passive case) and above ($\lambda_{exc}$ = 325 nm, active case) the ZnO bandgap. Figure 2 presents the PL spectrum of the ZnO NW (pink curve; more details in Figure S1) as well as of the CdSe/CdS NCs (black curve). The two laser lines used are also presented (green and red curves) along with the absorption spectrum of the NCs (blue curve). Both excitation lines as well as the ZnO emission overlap with the absorbance spectrum of the nanocrystals.

To prove the single photon emission, the antibunching behaviour of the photons emitted by the NC has been verified. In this view, the coincidence histogram *n(τ)* related to *I(τ)* has been collected, as described by the second order autocorrelation function:

$$g^{(2)}(\tau) = \frac{\langle :I(t+\tau)I(t): \rangle}{\langle I(t+\tau)\rangle\langle I(t)\rangle} \qquad (1)$$

where $I(t)$ is the emission intensity and $\tau$ is the delay between the two arms of an Hanbury-Brown and Twiss interferometer. The ':' represents the standard normal ordering in Glauber's formalism. We recall that if the $g^{(2)}$ function goes below 0.5, we can conclude that the nanocrystal is a single photon source. Experimentally, we used a pulsed laser ($\lambda_{exc}$ = 405 nm, pulse width = 50 ps) to measure the autocorrelation function with single photon counting detectors. Figure 3 reports the measured coincidence results for the nanocrystal shown in Figure 1-c. A value of $g^{(2)}(0)$ around 0.2 is obtained, by which we can infer that our hybrid device is truly made of a single nanowire coupled to a genuine single photon emitter. The fact that we do not get 0 for the autocorrelation function $g^{(2)}$ at zero delay is most likely due to spurious light coming from the system and the excitation laser or from the quality of the nanocrystal itself as was shown in Ref. 33.

The passive case is first investigated as ZnO NWs have been proven to behave as favourable nano-waveguides when coupled by light with energy below the energy bandgap[32,36]. We focus a 405 nm laser on one end of the NW (Figure 1-b) and evaluate whether the red-emitting NC can be excited on the opposite side (see schematic of Figure 1-a). Figure 4 gives evidence of the excitation of a single CdSe/CdS nanocrystal by the guided 405 nm laser propagating out of the NW. With our micro-PL set-up, we have the capability to pick up the spectrum at one place while exciting somewhere else at will. The white circle in inset of Figure 4 represents the spot where we collect the PL from the NC. These data show clearly that the nanocrystal is selectively excited only when the laser spot is on the nanowire (blue spot, on position). The emission of the single NC appears at 585 nm (blue curve) when the laser spot is on the nanowire (represented by the dotted rectangle) and disappears when the excitation lies outside the NW (red area). Effectively, we show here that some of the light from the excitation laser is coupled into the ZnO nanowire and then guided towards the other end of the NW. Then, some of the light comes off the end and excites the nanocrystal positioned just next to the nanowire (see Figure S4 for the FDTD simulation of the propagation). Upon comparing the PL intensity of the NC in this passive waveguiding configuration to the one obtained when the NC is directly excited by the 405 nm laser through our collecting microscope objective, we measure an excitation efficiency of 0.7 %. This value is remarkably high considering that the coupling of the light inside the nanowire is very inefficient at this stage. In fact, this value has to be corrected by two main factors: *i*- not all the incoming laser light is coupled into the NW and *ii*- the light in the NW gradually decreases during its propagation. For point *i*, we used a transmission experiment to find out that only 7 % of the incoming 405 nm laser is coupled into the NW by total internal reflection of the propagating laser light. According to the finite-difference time-domain (FDTD) simulation on the propagation of light inside the NW, only a small portion of the absorbed light is guided till the end (see Figure S4). For point *ii*, as the light propagates inside the NW, propagation losses have to be taken into account. There are due to light re-absorption by the NW itself and light scattering caused by surface roughness and defects. We can see indeed from Figure 2 that the 405 nm laser, even though below the zinc oxide bandgap, is still within the tail of the nanowire PL and re-absorption must occur. Hence, by taking into account the coupling losses only, we evaluate a passive ZnO waveguiding addressing efficiency of 10 % compared with the direct excitation scheme. Moreover, simulation in Figure S4 shows that for a nanowire of 280 nm diameter, our waveguide is slightly multimode and thus not optimised to match the radiation pattern of the emitter[37] (see also discussion on efficient addressing in SI). As a figure of merit (FOM), we use the addressing efficiency $\eta_a = 0.6 \ 10^{-4}$ (see supplementary information for the estimation of the NC addressing efficiency). Briefly, this FOM was obtained by careful estimation of the number of photons lost during the process of excitation then collection and finally detection of our set-up, for a given incident excitation power (thus number of photons created). We are positively confident that such pioneering demonstration of a hybrid NW-NC system can be much improved by optimising the nanowire growth quality as well as designing the right diameter for optimum optical mode propagation and excitation of the NC. Resonant excitation will also increase the effective absorption cross-section of our emitter and thus contributing to increasing the addressing efficiency. We can then use as a FOM the free space coupling efficiency defining in reference [27-29]. This would allow us to reach waveguiding addressing efficiencies similar or even higher than that of direct excitation (see supplementary information for more details of the estimation of the efficiency). We also demonstrated that emission coming from emitters directly excited, can emit back into the nanowaveguide (see Figure S8 in Supplementary information) with a fairly good efficiency (up to 20% measured by simple photoluminescence comparison), thus showing the feasibility for reversibility. These measurements were obtained by exciting directly a cluster of nanocrystals and collecting the light emitted from the other side of the nanowaveguide (see Figure S8).

The second case concerns the use of the NW as an active wavelength waveguide. When one end of the NW (Figure 1-b) is under the excitation of a 325 nm laser, a strong micro-PL at 377 nm can be observed in the region of the excitation spot, as well as on the opposite facet of the NW. In this case, the nanowire works as an active waveguide in a sense that the local ZnO PL coming up from the laser-excited area, acts as a light source inside the NW and is wave-guided to the opposite side of the nanowire. In order to investigate the PL propagation inside the NW, we also performed FDTD simulation of the electric field intensity distribution along the nano-waveguide (see Figure S5 in SI).

The PL of the NW shows a relatively broad spectrum around the ZnO bandgap energy (from 370 nm to 390 nm, see the spectrum in Figure 2). By spatially and spectrally analysing the emission along the entire nanowire length while exciting it on one end, we experimentally observe significant wavelength shifts and intensity absorption of the light propagating. Figure 5a shows three PLs collected respectively at the excitation spot (0 µm), at the middle length (4 µm) and at the output facet (7 µm) of the NW.

In the area of the excitation spot, the emitted localised NW PL shows a maximum at 377 nm with 12 nm of full-width-at-half-maximum. As the intrinsic light from ZnO propagates to the end of the NW, the emission shows an apparent peak shift to longer wavelengths. This is in fact not a shift but rather a stronger absorption at shorter wavelengths. This is due to a higher linear attenuation coefficient for the high energy side of the spectrum as it is closer to the intrinsic bandgap of the NW. To verify this, we plot the evolution of the PL intensity along the NW for two different wavelength emissions at 377 nm (close to the intrinsic ZnO bandgap) and 385 nm (away from the intrinsic ZnO bandgap) in Figure 5-b. Assuming that the absorption coefficient α is constant along the NW length and that scattered unguided light collected from the top is constant along the NW, we are able to fit the emitted PL intensities with an exponential decay as a function of the distance between the collection and the excitation spots. We find, indeed, that the absorption coefficient α is larger at 377 nm ($\alpha_{377}$ = 3 $10^3$ cm$^{-1}$) than at 385 nm ($\alpha_{385}$ = 2.1 $10^3$ cm$^{-1}$). These values are lower than the simulated ones using refractive index for bulk ZnO given by Postava *et al.*[31] (α = 8.4 $10^3$ cm$^{-1}$ at 377 nm and 6.9 $10^3$ cm$^{-1}$ at 385 nm). This discrepancy is expected as we are not dealing with bulk materials and one would need to take into account mode propagation as well as crystalline quality and surface defects for a characteristic wire-shaped nanomaterial. Nevertheless, the order of magnitude is the same. This mechanism of re-absorption has been studied before in the case of CdS NWs and mainly attributed to the so-called Urbach tail occurring at energies slightly below the band gap, and which can become dominant in 1D semiconductors due to electron-phonon coupling[38]. This is certainly a limiting factor of this type of waveguiding even though one can hope to engineer the growth of the nanowires so that this effect is lessened.

In the active configuration, similarly to the passive case, the single CdSe/CdS NC placed next to the facet of the NW (Figure 1-b), is locally excited by the NW-waveguided light output. Figure 6 presents the on- and off-spectrum of the NC when the laser spot is placed on or slightly off the NW, respectively.

To provide a quantitative comparison of the NC excitation efficiency triggered by ZnO in the active wavelength waveguiding configuration, we again compare it with respect to a direct free-space excitation like in the passive case. The NW-waveguided excitation produces a less intense emission than via direct excitation of the NC, and we find an efficiency of 1.1 % by direct spectrum comparison. Again, this efficiency can be much higher if we take into account the fact that spatial overlap between the NW diameter and the laser spot is only of 35 % (this is obtained by simply taking the ratio of the nanowire diameter/laser spot size, i.e. 280 nm/ 800 nm, both measured experimentally), thus only 35% at most of the incident light at 325 nm is absorbed by the nanowire as discussed in supplementary information. Under such conditions, we evaluate an addressing efficiency of 3.1 %. We find in this configuration that the addressing efficiency $\eta_a$ remains close to the value found for the passive case. One would expect to reach much better efficiencies by doing doping bandgap engineering of the nanowire material in order to shift the bandgap energy and the re-absorption energy range. Nevertheless, we clearly show that intrinsic photoluminescence from a single semiconductor nanowire is powerful enough to excite a single nanoemitter acting as a single photon source.

To conclude, we have proposed and demonstrated a photonic hybrid device where a single photon emitter is excited by light that is nano-waveguided by a ZnO semiconductor NW either below (passive case) or above (active case) its energy bandgap. In the research field of wavelength sensitive optical components for nanophotonic devices, the as-described ZnO NW - CdSe/CdS NC-based device places itself as a promising proof of principle. Encouraging excitation efficiencies of few percent through ZnO NW in the passive case have been achieved, with room to improve this efficiency and to potentially excite more effectively a nanoscale emitter than using standard microscopy techniques. In light of the intriguing and broad opto-electronic properties, size and shape dependent, exhibited by semiconductor NWs/NCs and abundantly reported in the literature, we envisage that similar photonic integration may involve different pairs of nano-actors and be extended to various application fields. For instance, instead of relatively weak PL at output facet, strong lasing effect might be generated through the use of NWs which will greatly increase the excitation efficiency on the single photon emitters[39–42]. Our presented technique is also limited in terms of yield of making such a system as the nanocrystals and the nanowires are randomly placed. Nevertheless, using manipulators, one could achieve systematic coupling[43]. We must stress that, even though the coupling of NCs and NWs together is random at the moment, when the elements coincide, we find that optical coupling occurs in 100% of the cases. A distance dependence emitter/nanowaveguide still remains to be done. We should also mention that using semiconductor nanowires can lead to electrical excitation through a voltage bias[44] and thus leading to electroluminescence coupling. By the same token, semiconductor NWs behave as promising sensitive photodetectors[45,46] and could be used in combination with our nanowire/nanocrystal structure for opening the route towards quantum nanophotonics and quantum nanodevices[47].

**Acknowledgement:** We thank E. Giacobino, S. Vezzoli and M. Sondermann for fruitful discussions. V. S. thanks C. Sartel for technical support. This work was supported by the Champagne-Ardenne region project NanoGain. We thank the regional platform "Nanomat" for the nanofabrication. This work was partially supported by C'nano IdF within the project SOPHOPOL. W. G. acknowledges the financial support of the China Scholarship Council (CSC). C. C. and A. B. would like to thank the CNRS programme PEPS for financial support and collaboration exchanges. L.C. acknowledges the MIUR-project MAAT "Molecular Nanotechnologies for Health and Environment". C. C. thanks the COST programme "Nonlinear Quantum Optics-NQO".

**Author Contributions:** W. G. and M. M. performed the experiment and analysed the data. W. G. fabricated the sample. N. R. and W. G. performed the FDTD simulation. V. S. provided the nanowire sample. M. V. and L. C. provided the nanocrystals. Q. G. analysed the data. A. B. and C. C. analysed the data and supervised the work. C. C. conceived the experiment. All authors contributed to the writing of the manuscript.

**Competing financial interests:**

The authors declare no competing financial interests.

18. Heiss, M. *et al.* Self-assembled quantum dots in a nanowire system for quantum photonics. *Nat. Mater.* **12,** 439–44 (2013).

19. Fedutik, Y., Temnov, V., Schöps, O., Woggon, U. & Artemyev, M. Exciton-Plasmon-Photon Conversion in Plasmonic Nanostructures. *Phys. Rev. Lett.* **99,** 136802 (2007).

20. Claudon, J. *et al.* A highly efficient single-photon source based on a quantum dot in a photonic nanowire. *Nat. Photonics* **4,** 174–177 (2010).

21. Gerhardt, I. *et al.* Strong Extinction of a Laser Beam by a Single Molecule. *Phys. Rev. Lett.* **98,** 033601 (2007).

22. Wrigge, G., Gerhardt, I., Hwang, J., Zumofen, G. & Sandoghdar, V. Efficient coupling of photons to a single molecule and the observation of its resonance fluorescence. *Nat. Phys.* **4,** 60–66 (2007).

23. Lee, K. G. *et al.* A planar dielectric antenna for directional single-photon emission and near-unity collection efficiency. *Nat. Photonics* **5,** 166–169 (2011).

24. Vamivakas, A. N. *et al.* Strong extinction of a far-field laser beam by a single quantum dot. *Nano Lett.* **7,** 2892–6 (2007).

25. Brokmann, X., Ehrensperger, M.-V., Hermier, J.-P., Triller, A. & Dahan, M. Orientational imaging and tracking of single CdSe nanocrystals by defocused microscopy. *Chem. Phys. Lett.* **406,** 210–214 (2005).

26. Aad, R. *et al.* Efficient Pump Photon Recycling via Gain-Assisted Waveguiding Energy Transfer. *ACS Photonics* **1,** 246–253 (2014).

27. Leuchs, G. & Sondermann, M. Light–matter interaction in free space. *J. Mod. Opt.* **60,** 36–42 (2012).

28. Sondermann, M., Lindlein, N. & Leuchs, G. Maximizing the electric field strength in the foci of high numerical aperture optics. *arXiv*:0811.2098v3 (2011).

29. Zumofen, G., Mojarad, N. M., Sandoghdar, V. & Agio, M. Perfect reflection of light by an oscillating dipole. *Phys. Rev. Lett.* **101,** 1–4 (2008).

30. Sallet, V. *et al.* Structural characterization of one-dimensional ZnO-based nanostructures grown by MOCVD. *Phys. Status Solidi* **247,** 1683–1686 (2010).

31. Postava, K. *et al.* Spectroscopic ellipsometry of epitaxial ZnO layer on sapphire substrate. *J. Appl. Phys.* **87,** 7820 (2000).

32. Voss, T. *et al.* High-order waveguide modes in ZnO nanowires. *Nano Lett.* **7,** 3675–80 (2007).
15

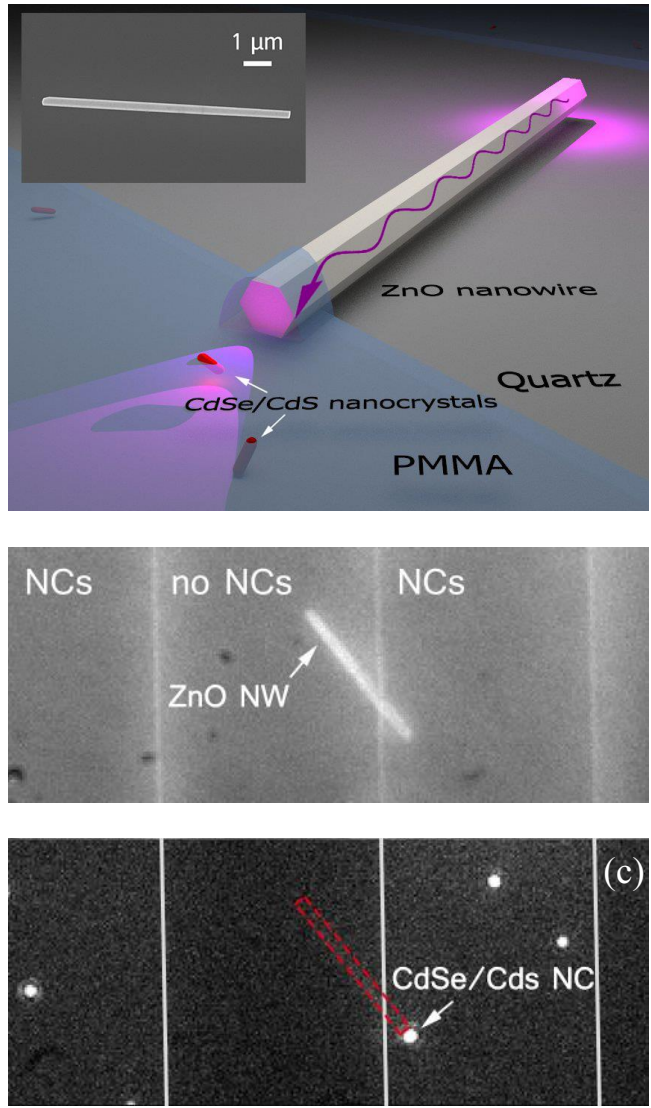

**Figure 1.**

**Single ZnO nanowire integrated with single CdSe/CdS nanocrystals. (a) Schematic of a ZnO nanowire crossing the nanocrystal and nanocrystal-free regions with representation of the excitation light on the upper end of the NW and waveguided light emission on the bottom end of the NW. Inset: SEM image of an individual ZnO nanowire. (b) Reflection image of the structures under the illumination of a 455 nm LED where a NW can be observed as well as the alternate regions with and without NCs; scale bar, 2 μm. (c) Scattered PL image obtained by shining with a 455 nm illumination, sufficient enough to excite single NCs, but below the bandgap of ZnO. A long-wavelength pass filter at 550 nm is used to select the emission of NCs only. One bright NC (bright dot) is found near the tip of the nanowire (represented in red dashed line); scale bar, 2 μm.**

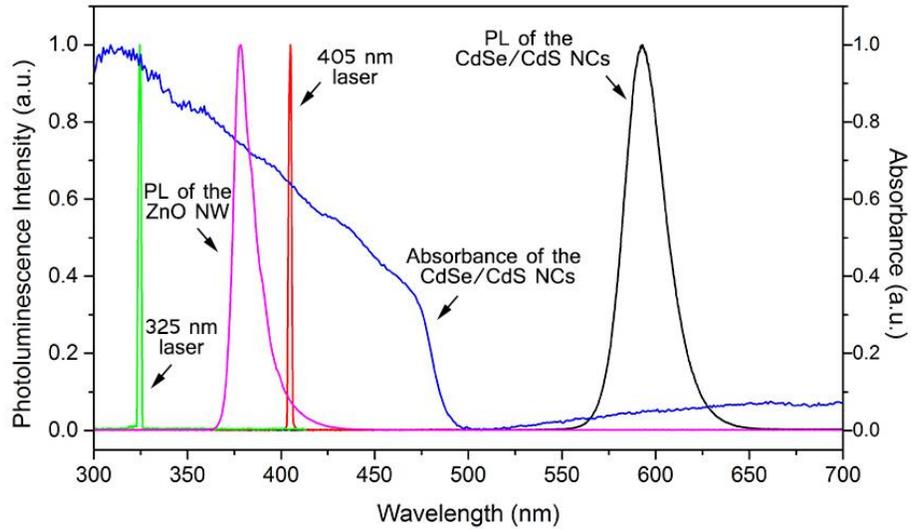

**Figure 2.**

**Laser lines, photoluminescence and absorbance spectra. The picture reports respectively the spectrum of the CdSe/CdS NCs emitting at 585 nm (black curve), absorbance spectrum of the CdSe/CdS NCs (blue curve, right axis), ZnO NW spectrum emitting at 377 nm (pink curve) and excitation laser lines at 325 nm (green curve) and 405 nm (red curve).**

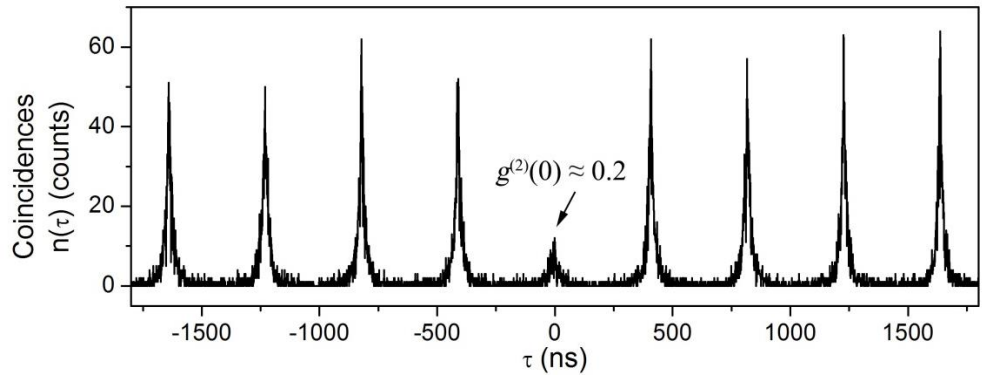

**Figure 3.**

**Coincidence histogram of the CdSe/CdS NC near the ZnO NW. A $g^{(2)}(0)$ less than 0.2 is measured, which indicates that this NC presents good antibunching behaviour.**

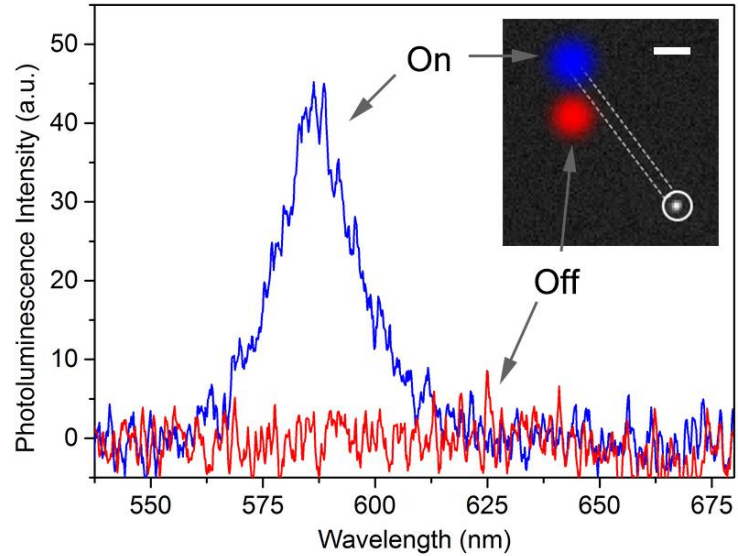

**Figure 4.**

**Single CdSe/CdS NC excited through a passive wavelength nano-waveguiding configuration. A long-wavelength pass filter cut-on 550 nm is used here to filter the 405 nm laser. The shape of the nanowire is drawn in white dashed line in inset. When the laser is focused in the on or off positon with respect to the end of the NW (blue and red spot in inset, respectively), the emission of the NC emerges (blue line) or disappears (red line) in the corresponding spectra; scale bar in the inset, 2 μm.**

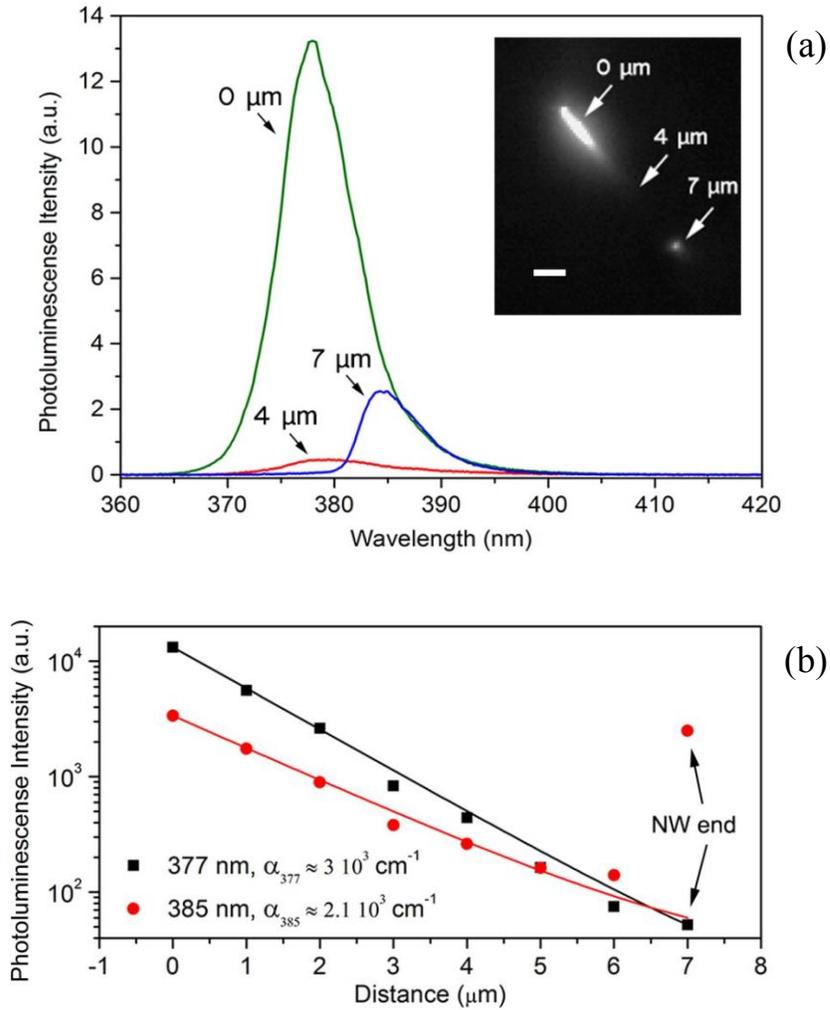

**Figure 5.**

**Propagation decay of the PL within the ZnO NW. (a) Spectra are taken at three different positions on the NW shown in the inset: 0 μm, 4 μm and 7 μm far from the excitation spot. A significant peak shift is observed because of the propagation losses; scale bar in the inset is 2 μm. (b) The evolution of intensity at 377 nm and 385 nm along the NW in red circle and black square, respectively. Exponential fits are plotted in solid lines for both wavelengths, which show that the decay rate, $α_λ$, is higher for 377 nm.**

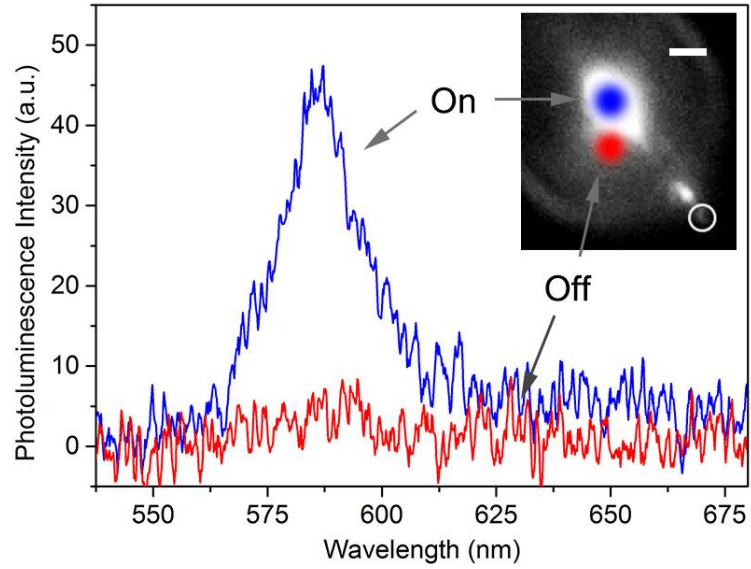

**Figure 6.**

**Single CdSe/CdS NC excited through an active wavelength nano-waveguiding configuration. In blue when the laser is on the NW and in red when it is far away from it. Part of the visible PL of the ZnO NW cannot be filtered by the long-wavelength pass filter cut-on 550 nm in the image in the inset, as it spectrally overlaps with the NC emission; scale bar in the inset, 2 μm.**